\documentclass[12pt]{article}
\usepackage{geometry}
\usepackage{makeidx}
\usepackage[T1]{fontenc}
\usepackage[dvipsnames,svgnames,table]{xcolor}
\usepackage{epstopdf}
\usepackage{epsfig}
\usepackage{textcomp}
\usepackage{hyperref}
\usepackage{amsmath}
\usepackage{amssymb}
\usepackage{multirow}
\usepackage{multicol}
\usepackage{booktabs}
\usepackage{lastpage}
\usepackage{hyperref} 
\usepackage{graphicx}
\usepackage{enumerate}
\usepackage{threeparttable}
\usepackage{float}
\usepackage{array}
\usepackage{cite}
\usepackage{color,soul}
\usepackage{listings}
\usepackage{tikz}
\usepackage{longtable}
\makeatletter
\renewcommand\section{\@startsection{section}{1}{\z@}%
{-2.5ex \@plus -1ex \@minus -.2ex}%
{2.3ex \@plus.2ex}%
{\normalfont\large\bfseries}}
\renewcommand\subsection{\@startsection{subsection}{1}{\z@}%
{-2.5ex \@plus -1ex \@minus -.2ex}%
{2.3ex \@plus.2ex}%
{\small\bfseries}}

\begin{document}
\bigskip
\title{\textbf{Unveiling secondary particle generation in a SONTRAC detector through GEANT4 simulations}}
\medskip
\author{\small A. Ilker Topuz$^{1}$}
\medskip
\date{\small$^1$Manipal Centre for Natural Sciences, Centre of Excellence, Manipal Academy of Higher Education, Manipal 576104, India\\aitopuz@protonmail.com}
\maketitle
\begin{abstract}
SOlar Neutron TRACking (SONTRAC) is a detector concept based on a bundle of plastic scintillators by aiming at tracking the solar neutrons through the generation of the secondary particles such as protons from the (n, np) and (n, p) processes. In this study, in addition to the particle population, the energy spectra of the secondary particles including protons, gamma rays, electrons, alphas, and ions that are produced either due to the interaction between the fast neutrons and a SONTRAC detector or through the interplay between the secondary particles and the detector components are determined by means of GEANT4 simulations. The detector geometry in the present study consists of 34$\times$34 Kuraray Y11-200(M) fibers, the composition of which includes polystyrene for the fiber core, poly(methyl methacrylate) (PMMA) for the first clad, and fluorinated PMMA for the second clad. The current fiber bundle is irradiated with a planar vertical neutron beam of 0.2$\times$0.2 cm$^{2}$ by using an energy list composed of 20, 40, 60, 80, and 100 MeV where the number of incident neutrons is $10^5$, and it is first revealed that a non-negligible number of secondary protons are generated by the fast neutron bombardment; however, the population of these secondary protons is still low compared to the incident beam, i.e. in the order of $10^3$. Secondly, it is also observed that the energy spectrum of secondary protons exhibits a decreasing trend that is limited by the kinetic energy of incident neutrons. Additionally, the range of the secondary protons along with the deposited energy is computed, and it is demonstrated that a significant portion of the generated protons lose their entire energy and stop within the present SONTRAC detector. Finally, a 34$\times$34 pixel grid detector is introduced on each side of the fiber bundle to collect the optical photons produced from the energy deposition in the scintillation fibers, and the trajectory of the secondary protons on the pixel grid is shown by using a fast neutron beam of 100 MeV.
\end{abstract}
\section{Introduction}
\label{Intro}
The solar flares might lead to the generation of the fast neutrons along with the high-energetic gamma rays in the energy level of MeV~\cite{ligenfelter1965high,daniel1969search,murphy1984solar,ramaty1987nuclear,mitchell2022physics}, and this population of the fast neutrons generally has a significance in the space applications such as space dosimetry~\cite{cataldo2020neutron,sato2020recent,sihver2021space,roffe2021neutron}. Among the existing detector systems for the fast neutron tracking~\cite{yamaoka2021solar, muraki2023detection, Yamaoka:2023O4} is the SOlar Neutron TRACking (SONTRAC) detector concept~\cite{de2019solar,mitchell2021development,de2023next}, which is based on a fiber bundle with an array of plastic scintillators. In this kind of detection concept, the fast neutrons are tracked through the emission of the secondary protons from the (n, np) and (n, p) processes, the energy deposition of which yields the optical photons in the scintillation fibers. Thus, the present study is devoted to the determination of the energy spectrum besides the population for the secondary particles due to the fast neutron irradiation through the GEANT4 simulations~\cite{agostinelli2003geant4, TopuzGithubSONTRAC}. Regarding the current simulation setup, a 34$\times$34 fiber bundle made out of the Kuraray Y11-200(M) scintillation fibers~\cite{Kuraray} is defined. An incident planar vertical monoenergetic neutron beam of 20, 40, 60, 80, and 100 MeV with an initial population of $10^5$ neutrons is utilized. In addition to the population of the secondary particles, the energy spectrum for the secondary protons, gamma rays, electrons, alphas, and ions is determined. Furthermore, the deposited energy by the secondary protons together with the range within the present SONTRAC detector is provided. Into the bargain, a 34$\times$34 pixel grid detector on every side of the fiber bundle is introduced in order to collect the optical photons. This study is organized as follows. In section~\ref{Setup}, the current SONTRAC detector features accompanied by the simulation properties in GEANT4 are stated, while the simulation outcomes related to the secondary particles are exhibited in section~\ref{Outcomes}. Finally, section~\ref{Conclusion} incorporates the conclusions drawn from the current GEANT4 simulations.
\section{Simulation setup}
\label{Setup}
\begin{figure}[H]
\begin{center}
\includegraphics[width=8.9cm]{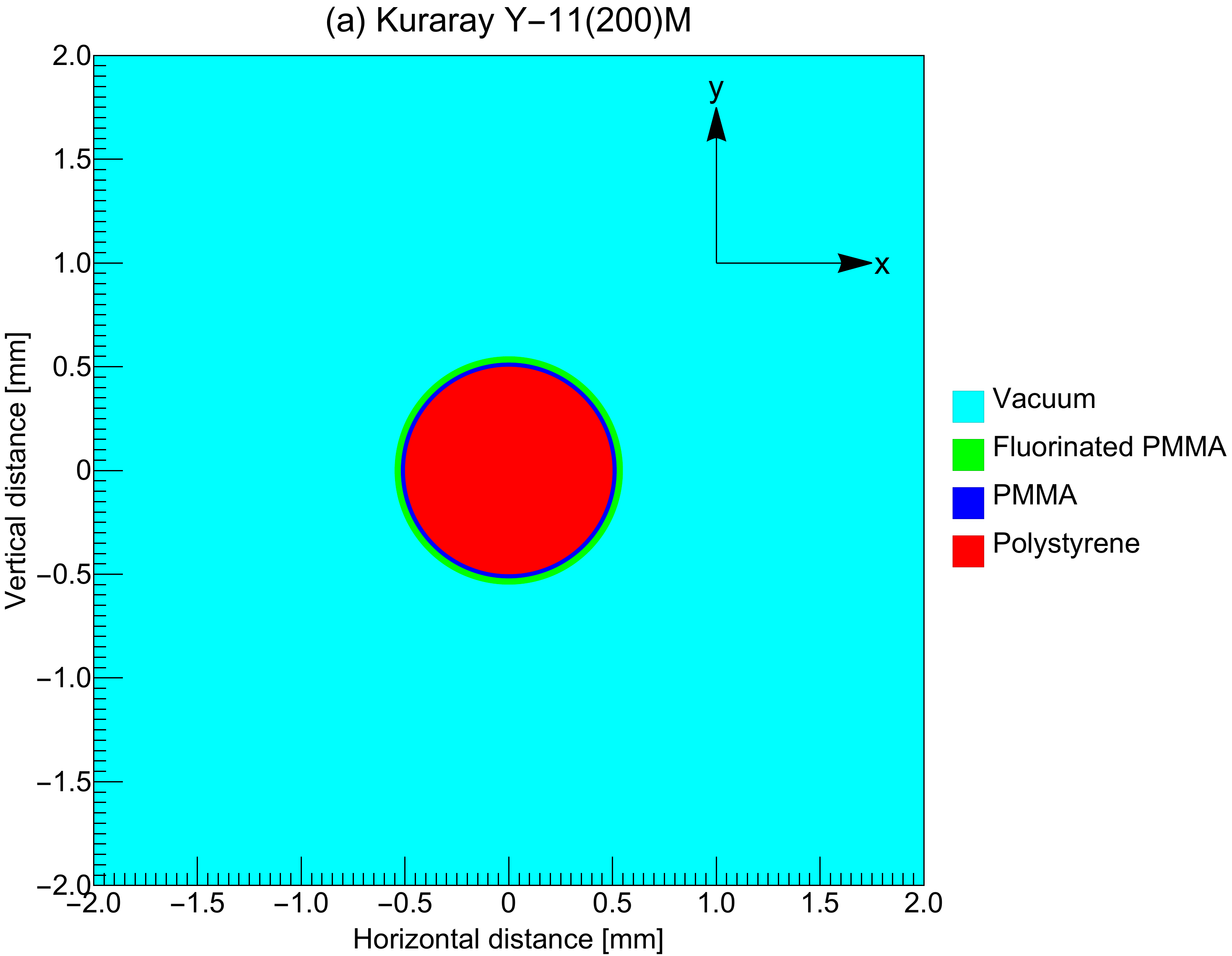}
\includegraphics[width=7cm]{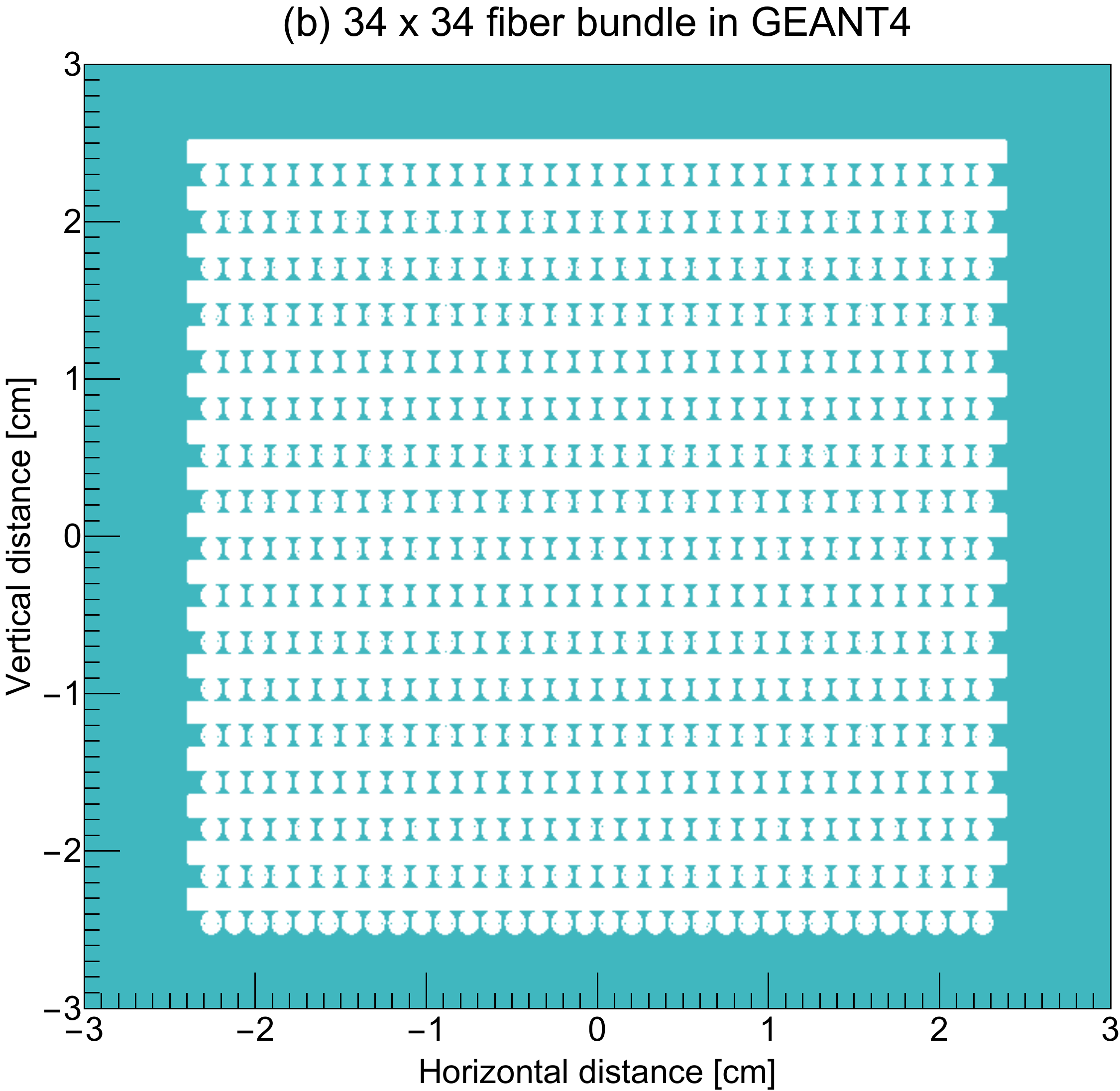}
\caption{Geometry of SONTRAC detector based on Kuraray Y11-200(M) fibers in GEANT4 simulations: (a) Kuraray Y11-200(M) fiber dimension and (b) 34$\times$34 fiber bundle.}
\label{Fiber_geometry}
\end{center}
\end{figure}

As previously mentioned, the SONTRAC detector in the present study is assembled from the Kuraray Y11-200(M) scintillation fibers, and each of these scintillation fibers is considered as a multi-layer co-centered cylindrical medium where the fiber core constituting the first layer is manufactured from polystyrene, while the second and third layers are the cladding materials, the chemical compositions of which are poly(methyl methacrylate) (PMMA) and fluorinated PMMA, respectively. Regarding the layer dimensions, the radius of the fiber core is 0.5 mm, whereas the radius of the first cover is 0.52 mm, and the radius of the second cover is 0.55 mm as described in Fig.~\ref{Fiber_geometry}(a). The length of each Kuraray Y11-200(M) fiber is around 5 cm, and the pitch size is 1.36 mm.
As shown in Fig.~\ref{Fiber_geometry}(b), a bundle of 34$\times$34 fibers is built along the x-direction as well as the z-direction. The surrounding medium in the current GEANT4 simulations is vacuum. The GEANT4/NIST database is followed in the definition of every material. A planar vertical beam of 0.2$\times$0.2 cm$^{2}$ along the y-direction is used, and the incident beam consists of 10$^{5}$ neutrons by utilizing a constant kinetic energy of 20, 40, 60, 80, and 100 MeV. The reference physics list preferred in these simulations is FTFP$\_$BERT$\_$HP. The secondary particle tracking inside the SONTRAC detector is performed by G4Step, and the generated secondary particles including protons, gamma rays, electrons, alphas, and ions are analyzed through Python by taking all the existing detector components into consideration.

\begin{figure}[H]
\begin{center}
\includegraphics[width=8.9cm]{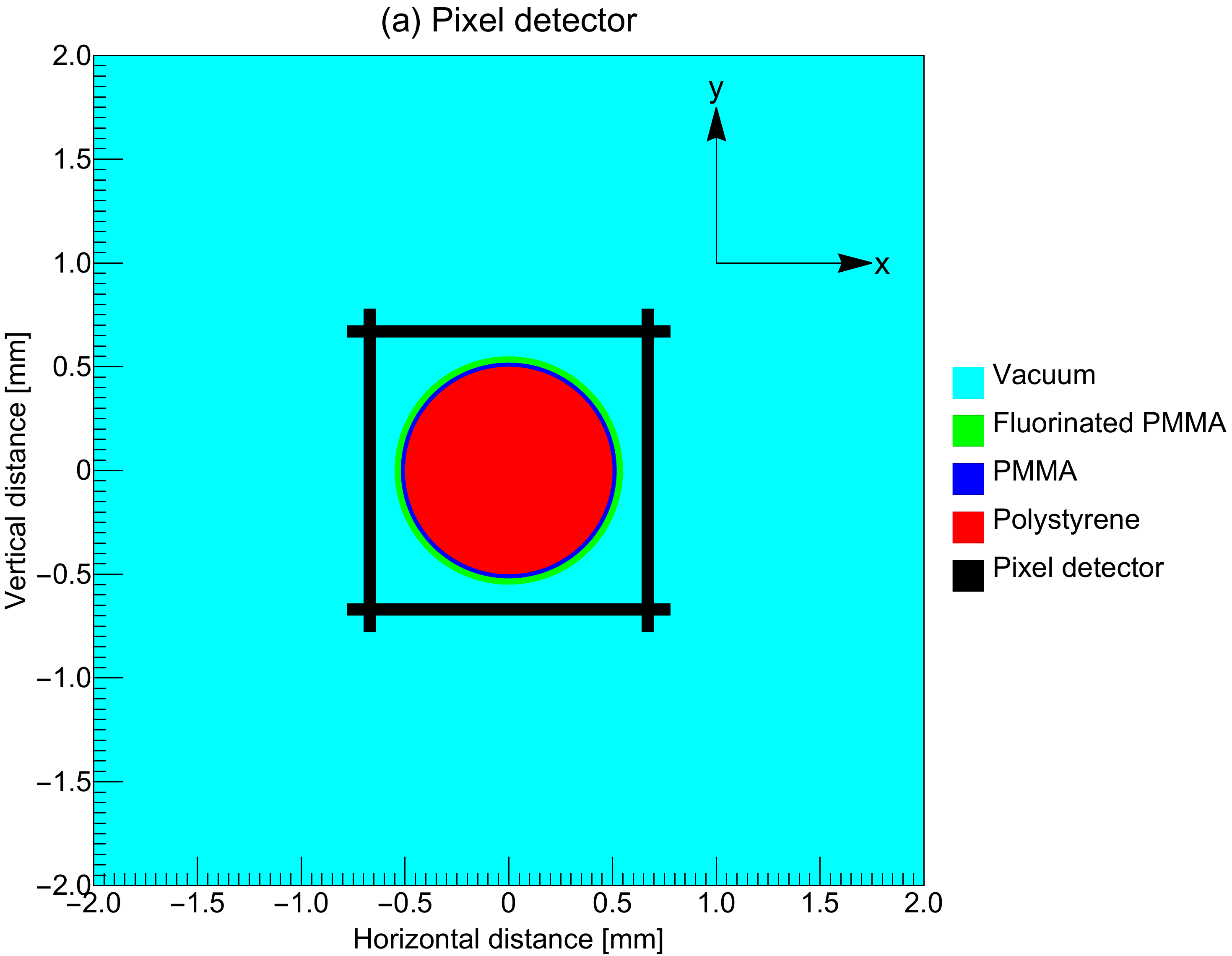}
\includegraphics[width=7cm]{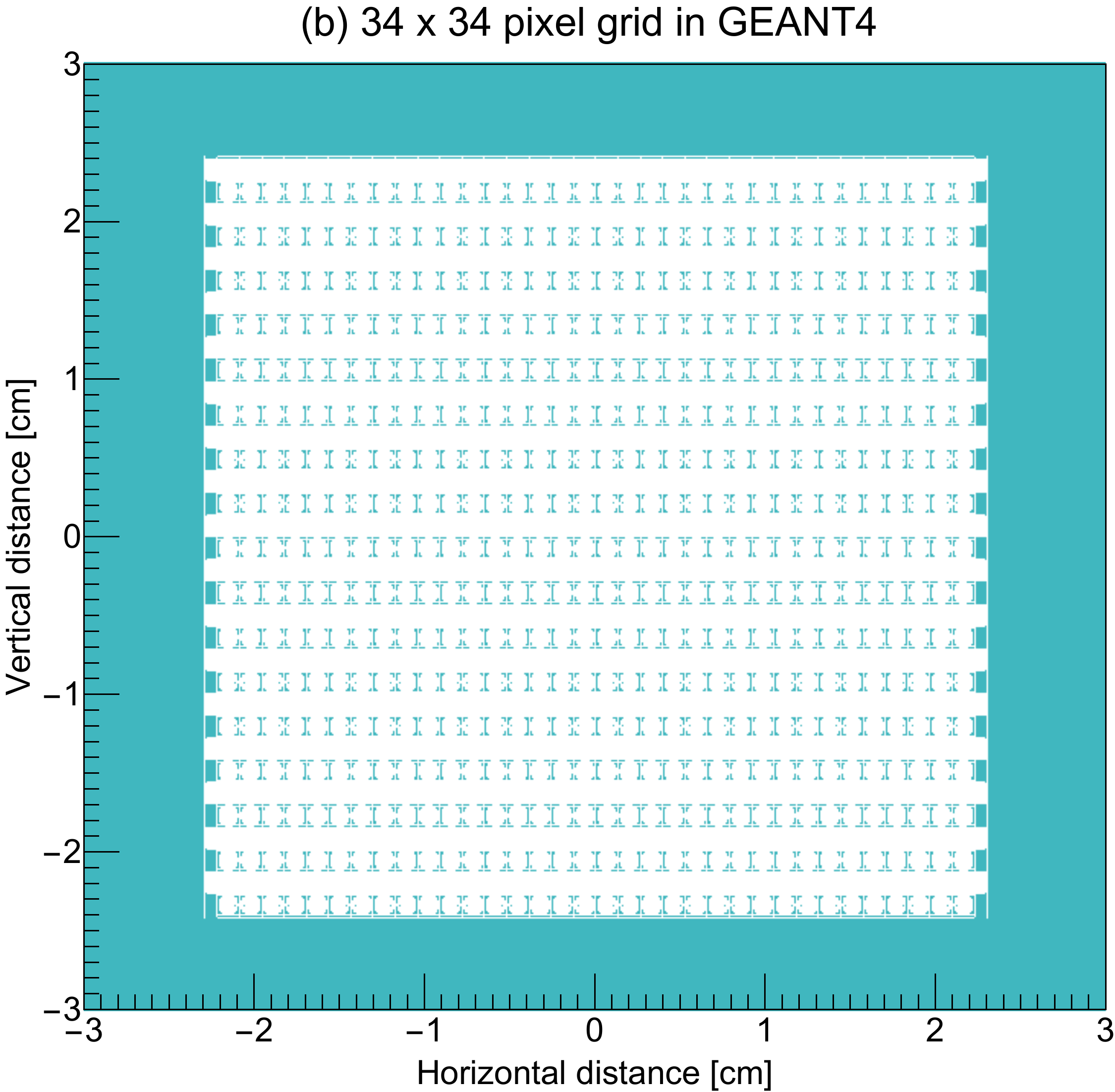}
\caption{Introduction of pixel detector in GEANT4 simulations: (a) Square pixel of 1.36$\times$1.36 mm$^{2}$ and (b) 34$\times$34 pixel grid.}
\label{Pixel_geometry}
\end{center}
\end{figure}

On top of the secondary particle tracking, the generation of the optical photons from the scintillation fibers owing to the energy deposition of the secondary protons is also performed. The refractive indices for the fiber components that include the fiber core, the first clad, and the second clad are 1.59, 1.49, and 1.42, respectively. By using an equal production rate, the energies of the optical photons are 1.75 and 2 eV, and the absorption length for the fiber elements is selected at 1 m in order to enhance the computation speed. Additionally, the scintillation yield is 10 per keV, the resolution scale is 1, and the scintillation time constant is 10 ns. Furthermore, a Birks constant of 0.126 mm per MeV is applied to the Kuraray Y11-200(M) fiber core. For the release of the optical photons, EmStandardPhysics$\_$option4 is included in addition to FTFP$\_$BERT$\_$HP. In the optical photon simulations, the number of incident neutrons is 100 to facilitate the demonstration. A 34$\times$34 pixel grid is introduced in order to gather the optical photons as shown in Fig.~\ref{Pixel_geometry}(a), and the area of each pixel pointing out each separate Kuraray Y11-200(M) scintillation fiber is 1.36$\times$1.36 mm$^{2}$. The pixel grid has a thickness of 0.0236 mm and is situated approximately 0.1 mm away from the fiber bundle.
The SONTRAC fiber bundle is surrounded by inserting a pixel grid on each side as a means to detect the generated optical photons from every face of the fiber bundle along the x-coordinate as well as the z-coordinate, i.e four pixel grids in total. Finally, a pixel grid of 34$\times$34 on one side of the fiber bundle in GEANT4 is shown in Fig.~\ref{Pixel_geometry}(b). 
\section{Simulation outcomes}
\label{Outcomes}
\subsection{Population and energy spectrum of secondary particles}
As stated beforehand, the solar neutron tracking via the SONTRAC detector is founded on the generation of the secondary protons, the energy deposition of which occurs in the scintillation fibers by emitting a number of optical lights. Thus, the population of the secondary protons as well as the corresponding energy spectrum constitutes a fundamental aspect of the solar neutron detection in the case of the SONTRAC detector. 
\begin{figure}[H]
\begin{center}
\includegraphics[width=10cm]{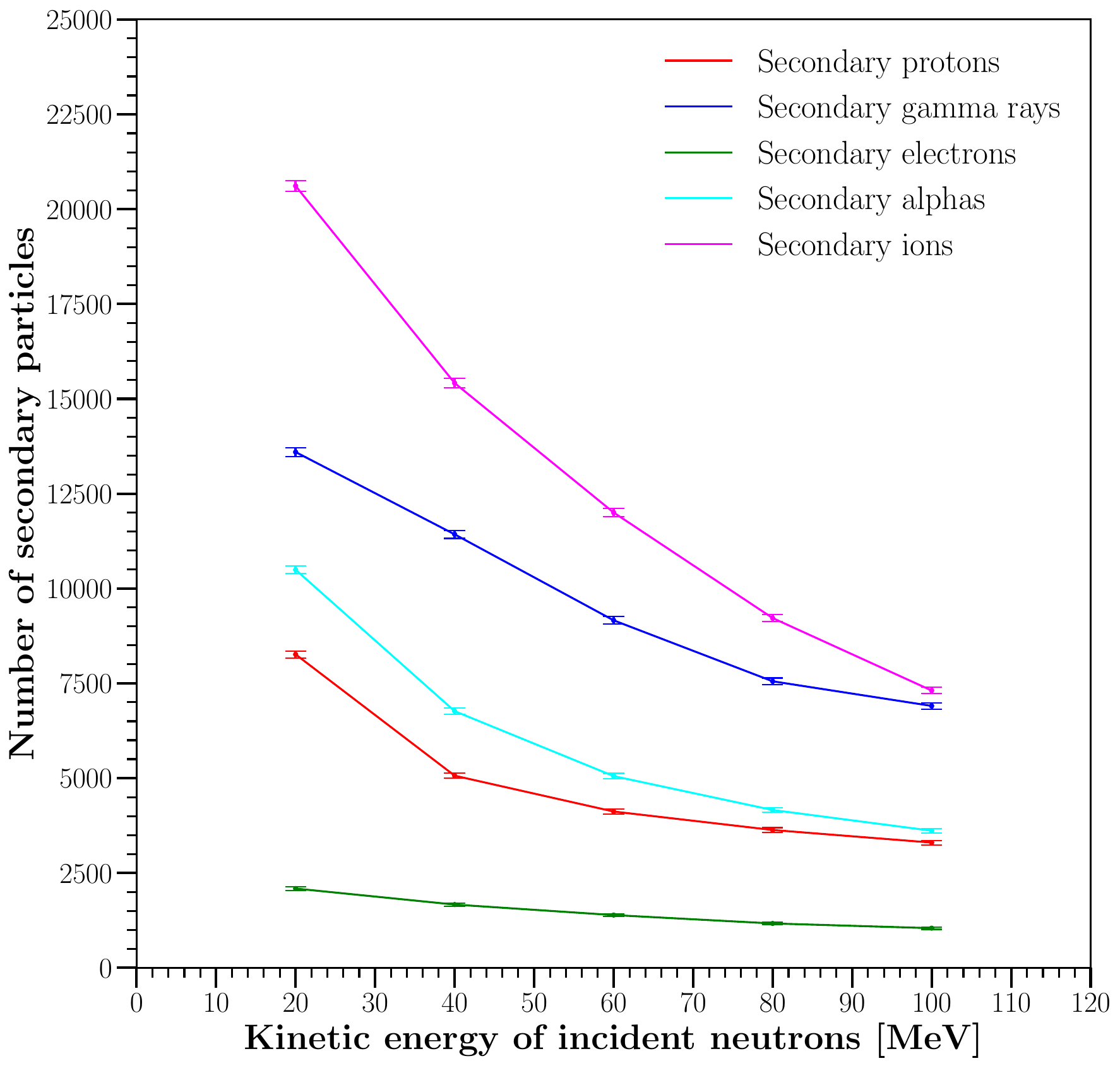}
\caption{Number of secondary particles within the present SONTRAC detector irradiated by fast neutron beam including protons, gamma rays, electrons, alphas, and ions.}
\label{Secondary_population}
\end{center}
\end{figure}
When the current simulation outcomes are investigated, the present GEANT4 simulations show that the secondary proton population is low in comparison with the incident neutron beam consisting of 10$^{5}$ neutrons, i.e. in the order of 10$^{3}$, as indicated in Fig.~\ref{Secondary_population}. This also means that only a small portion of the incoming fast neutrons might be detected within the current SONTRAC detector by means of secondary protons; on the other hand, it should be noted that a certain number of incident neutrons do not hit any scintillation fiber for the reason of the pitch size. Secondly, the number of secondary gamma rays that are mainly produced through the (n, n$\gamma$) process is determined as depicted in Fig.~\ref{Secondary_population}, and it is seen that the number of these secondary gamma rays is approximately two times greater than that of secondary protons in the energy interval between 40 and 100 MeV. Subsequently, Fig.~\ref{Secondary_population} indicates the population of the secondary electrons, and it is revealed that the number of secondary electrons is as low as that of secondary protons. It is worth mentioning that the source of secondary electrons regularly is the Compton process of secondary gamma rays. Despite their very short range within the detector setup,  the number of secondary alphas as well as the number of secondary ions is also obtained as illustrated in Fig.~\ref{Secondary_population}. 
\begin{figure}[H]
\begin{center}
\includegraphics[width=7.5cm]{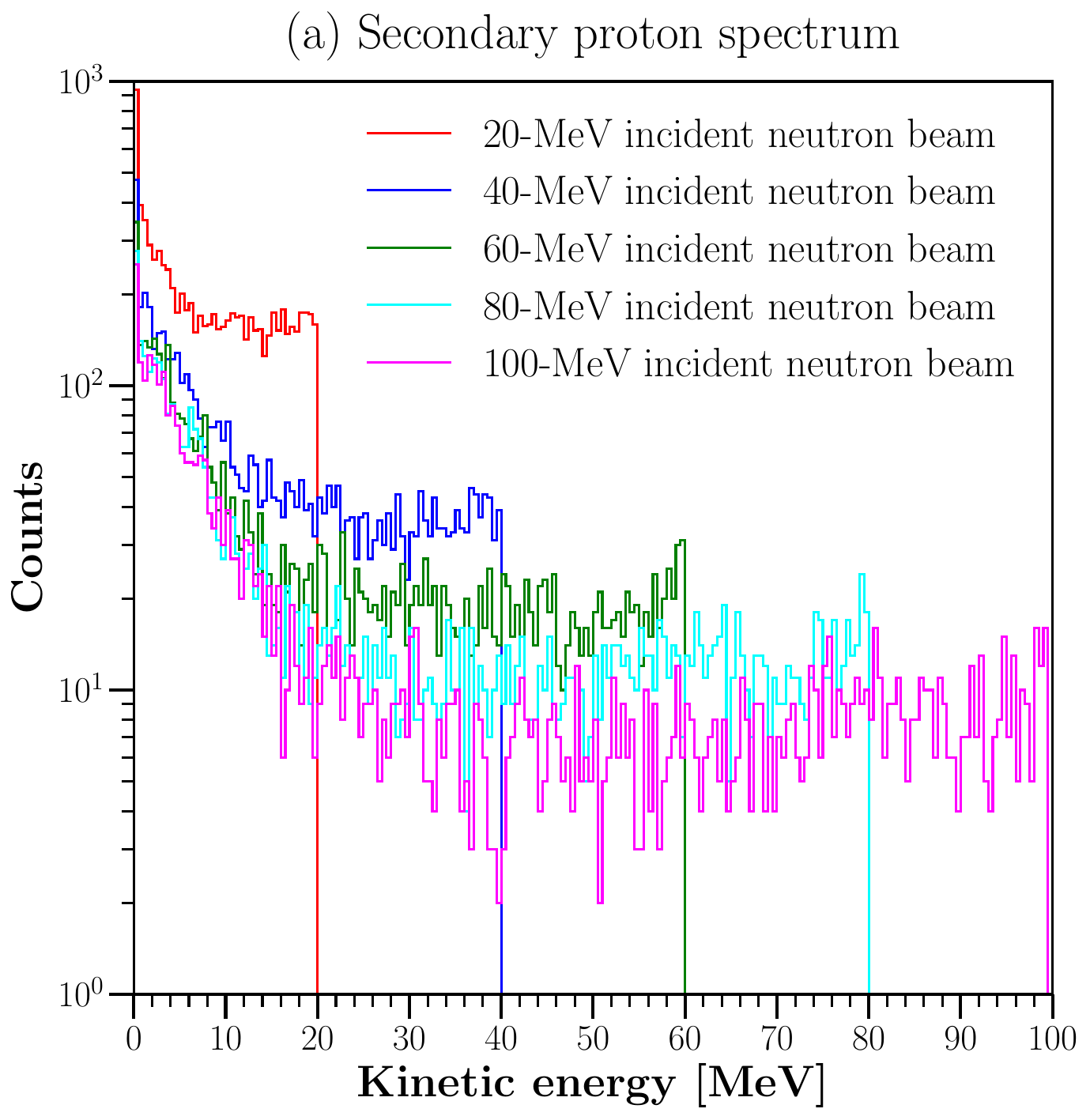}
\includegraphics[width=7.5cm]{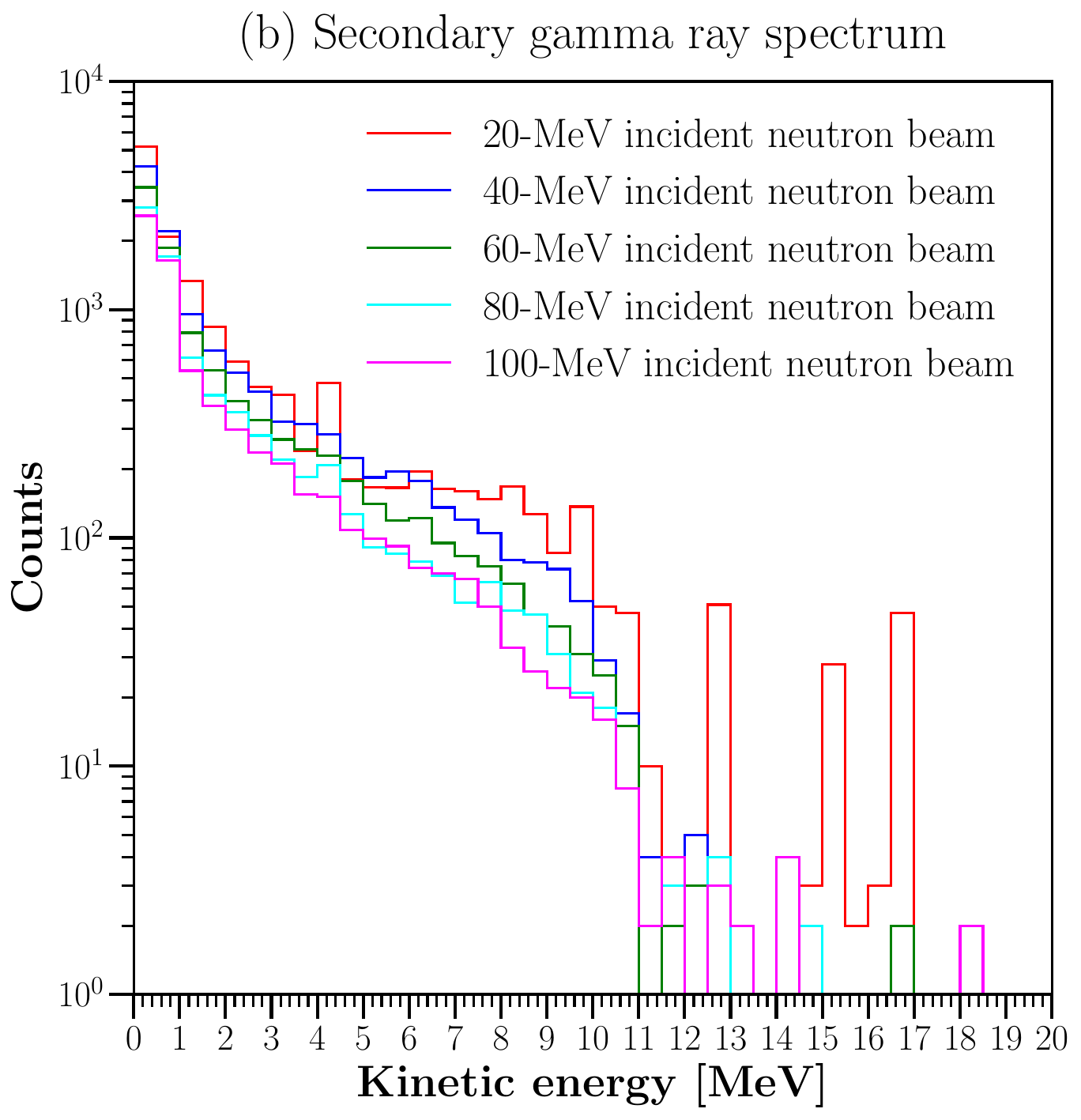}
\includegraphics[width=7.5cm]{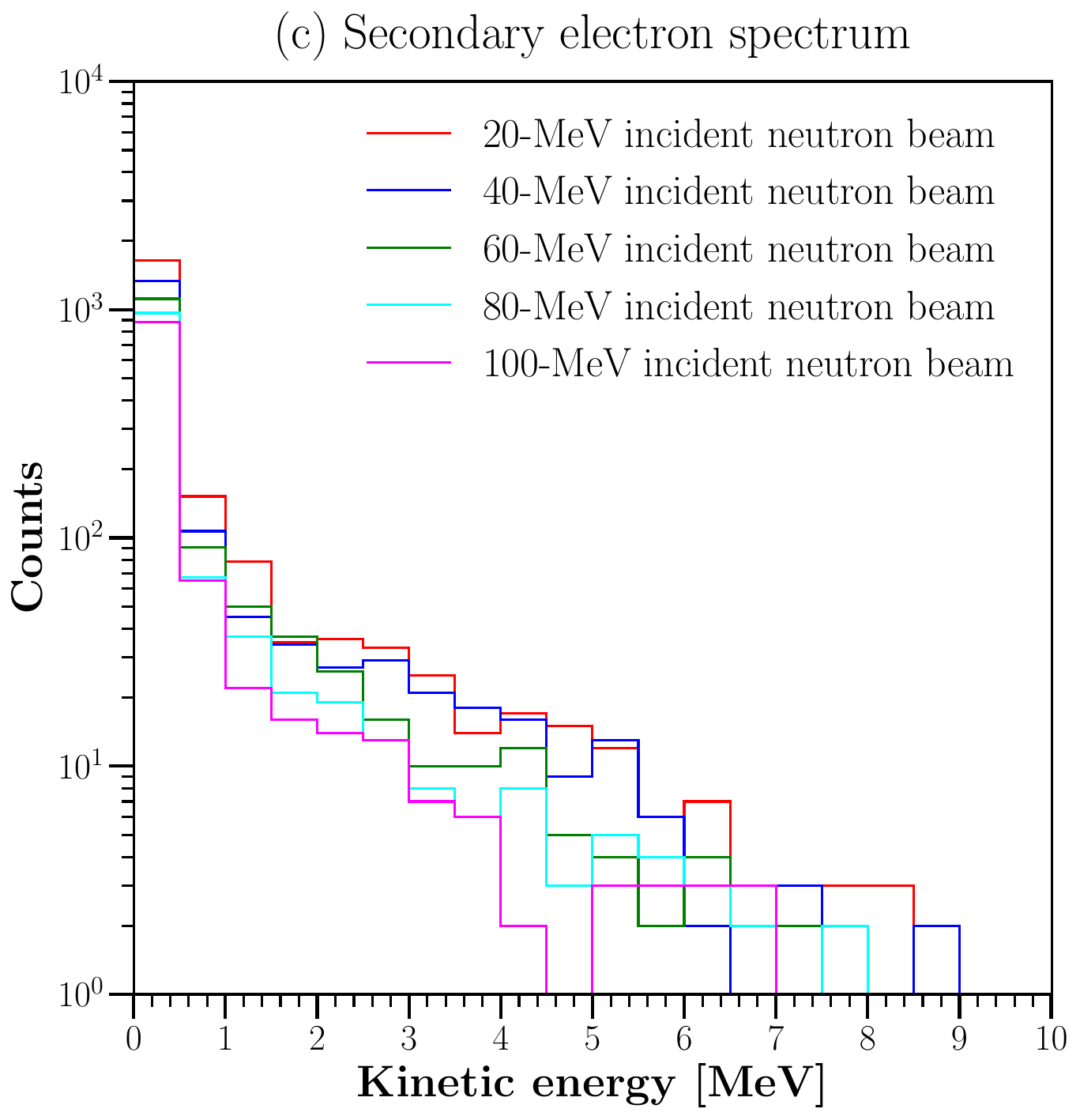}
\includegraphics[width=7.5cm]{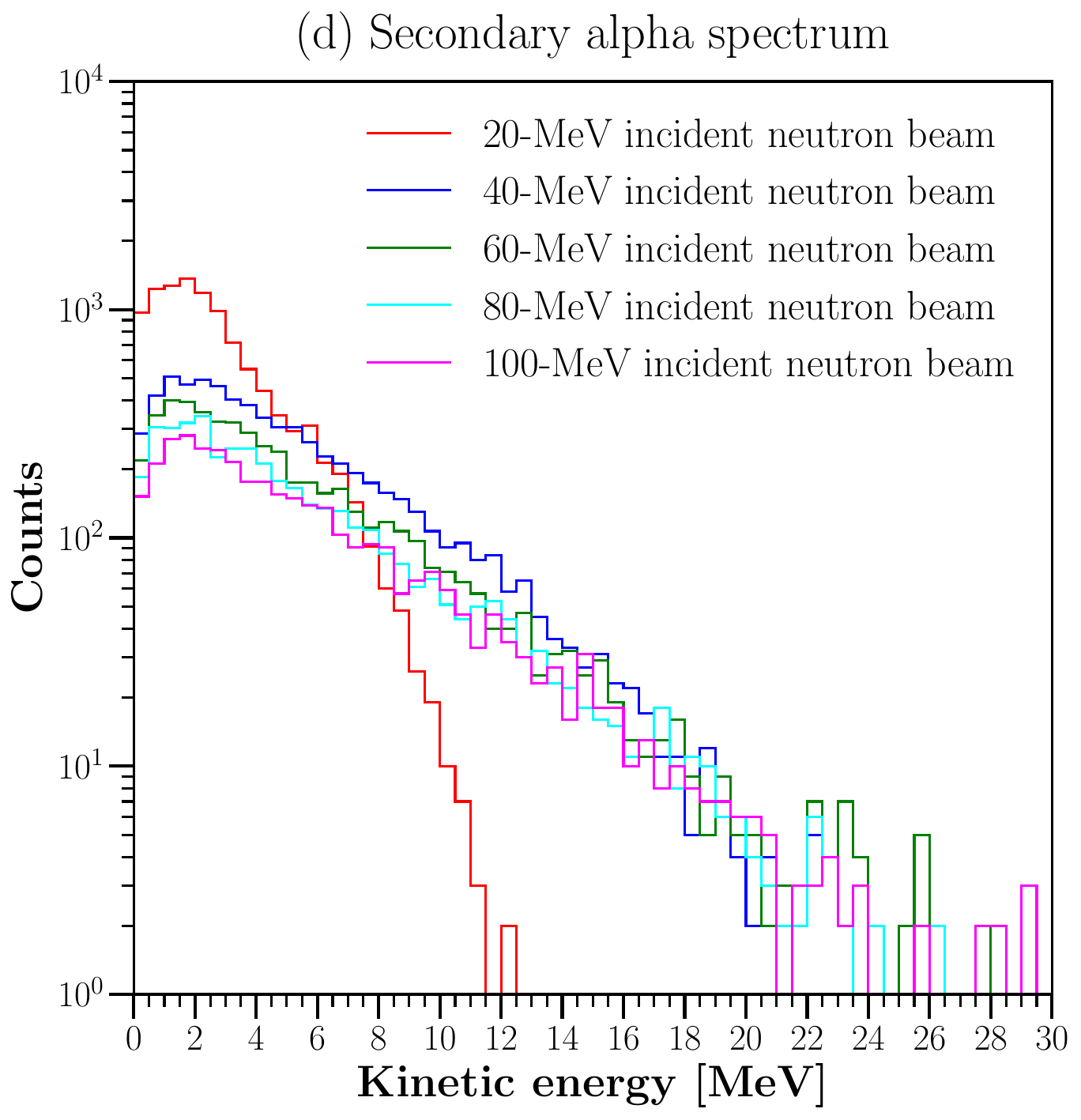}
\includegraphics[width=7.5cm]{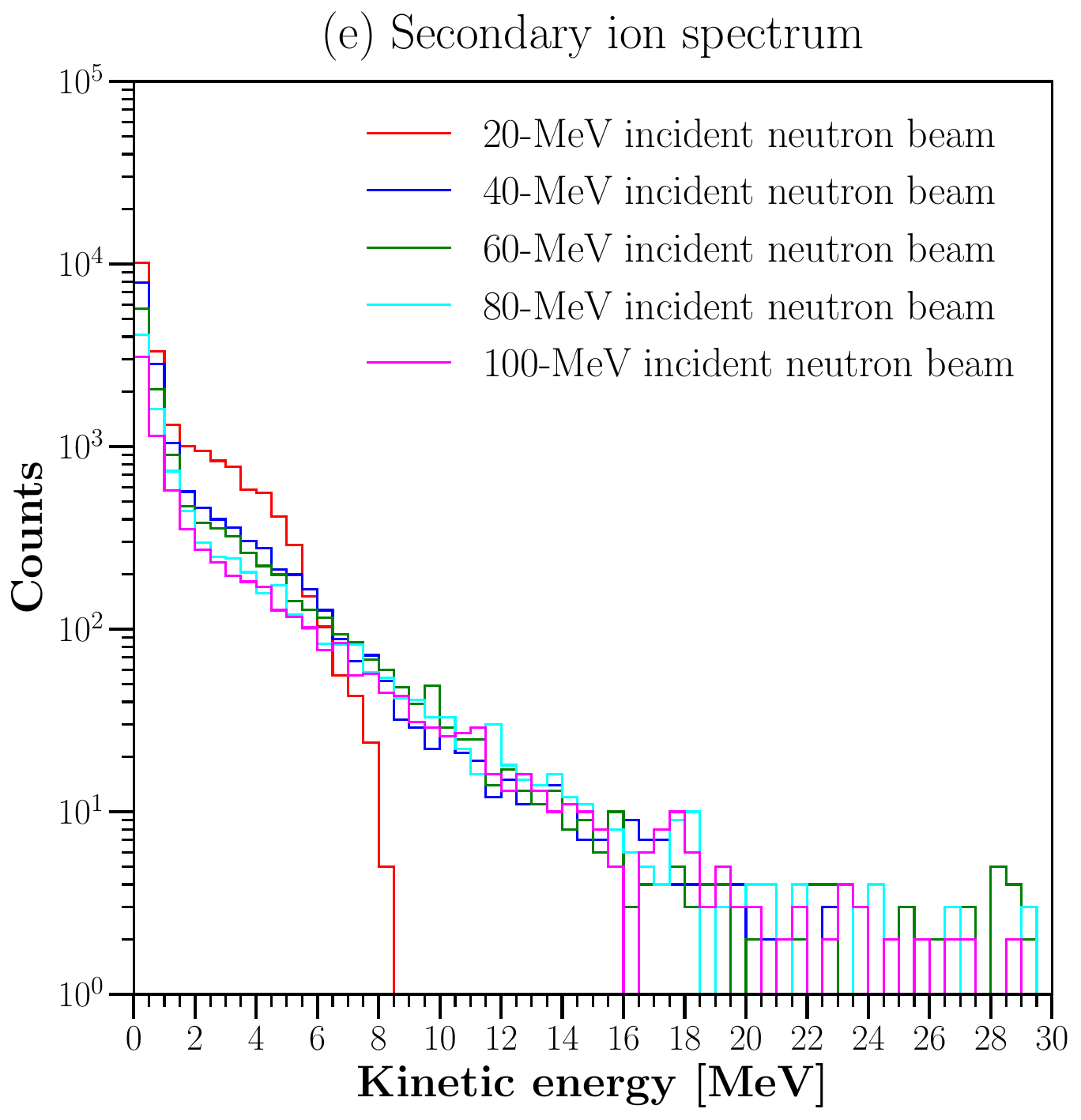}
\caption{Energy spectrum of secondary particles with a step length of 0.5 MeV within the present SONTRAC detector irradiated by fast neutron beam: (a) protons, (b) gamma rays, (c) electrons, (d) alphas, and (e) ions.}
\label{Secondary_energy_spectrum}
\end{center}
\end{figure}
The current simulation results demonstrate that the populations of the secondary alphas and the secondary ions together with the secondary gamma rays comprise the majority of the secondary particles due to the fast neutron irradiation. Without any exception, it is concluded from Fig.~\ref{Secondary_population} that the number of secondary particles is reduced when the kinetic energy of the incident neutron beam is increased.

In the next step, the energy spectra of the secondary particles owing to the fast neutron beam are determined as shown in Figs.~\ref{Secondary_energy_spectrum}(a)-(e), and it is noticed that the energy spectrum of the secondary protons is bounded by the kinetic energy of the incident neutrons, which means that the maximum kinetic energy of the secondary protons is approximately equal to that of the incident neutrons, and the energy interval of the secondary protons is significantly larger compared to that of the remaining secondary particles. Additionally, the kinetic energy of the generated protons is generally above the energy cut-off for the generation of optical photons, which is usually in the keV level. The trends in the energy spectra of the secondary gamma rays and the secondary electrons roughly resemble as seen in Figs.~\ref{Secondary_energy_spectrum}(b)-(c), whereas the spectrum shapes of the secondary alphas and the secondary ions are coarsely similar with the exception of the decrement rate as shown in Figs.~\ref{Secondary_energy_spectrum}(d)-(e). In every energy spectrum, the population of the low-energetic particles constitutes the highest portion, and the number of secondary particles is reduced when the kinetic energy of the secondary particles is augmented.
\subsection{Range and energy deposition of secondary protons}
\begin{figure}[H]
\begin{center}
\includegraphics[width=8cm]{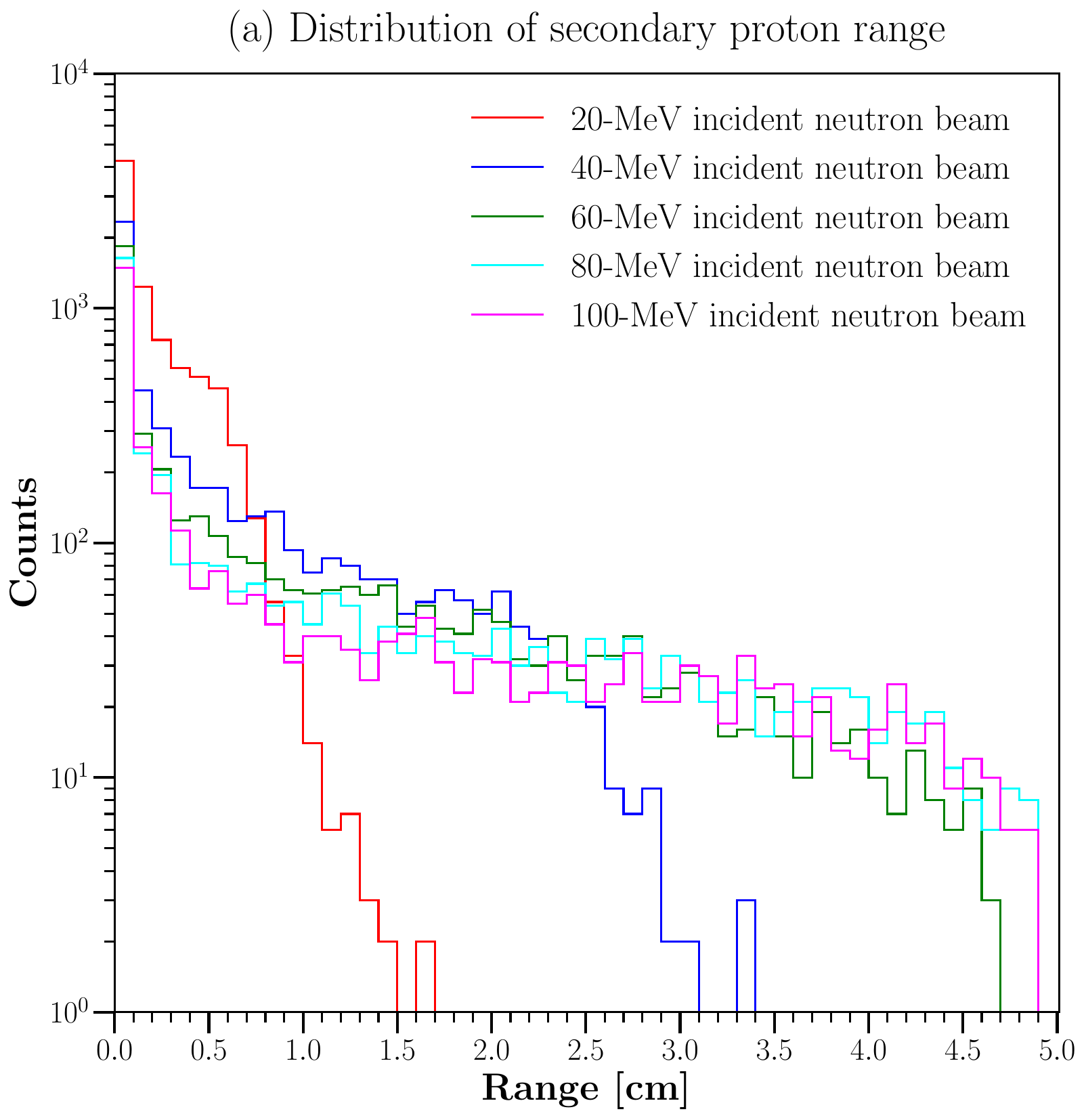}
\includegraphics[width=8cm]{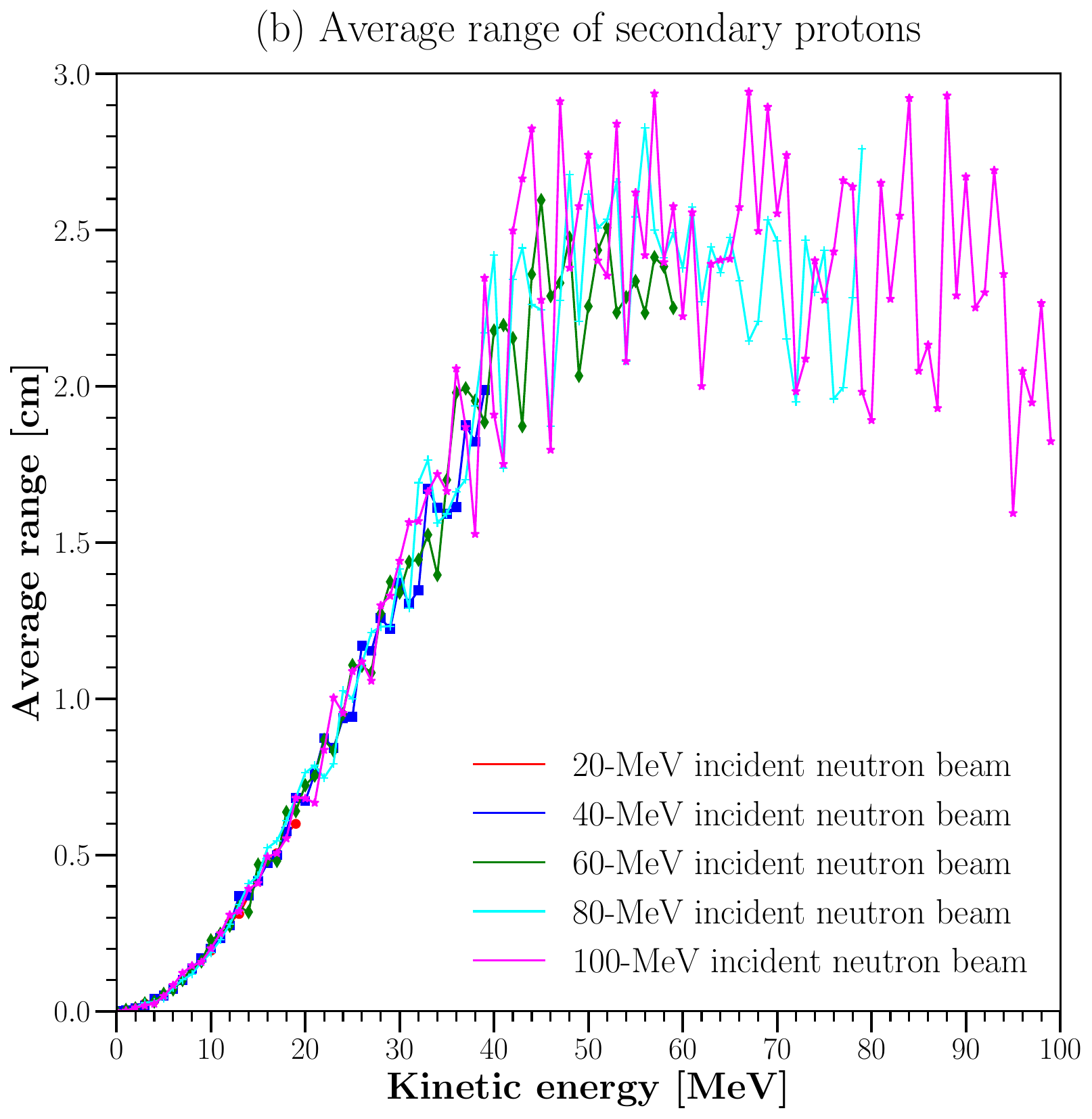}
\caption{Range of secondary protons within the present SONTRAC detector irradiated by fast neutron beam: (a) distribution of secondary proton range with a step length of 0.1 cm and (b) average proton range with a bin size of 1 MeV.}
\label{Proton_range}
\end{center}
\end{figure}

\begin{figure}[H]
\begin{center}
\includegraphics[width=8cm]{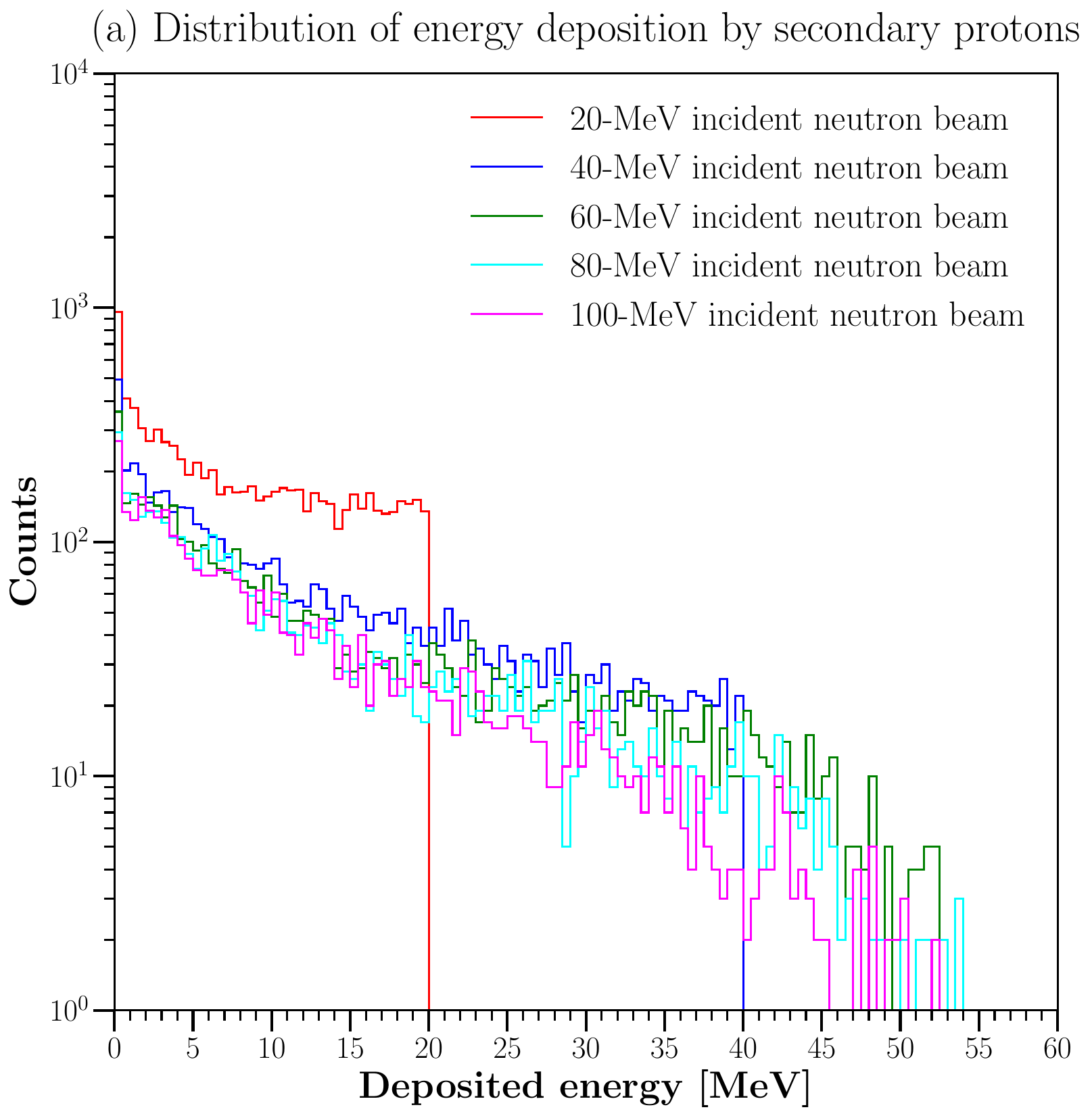}
\includegraphics[width=8cm]{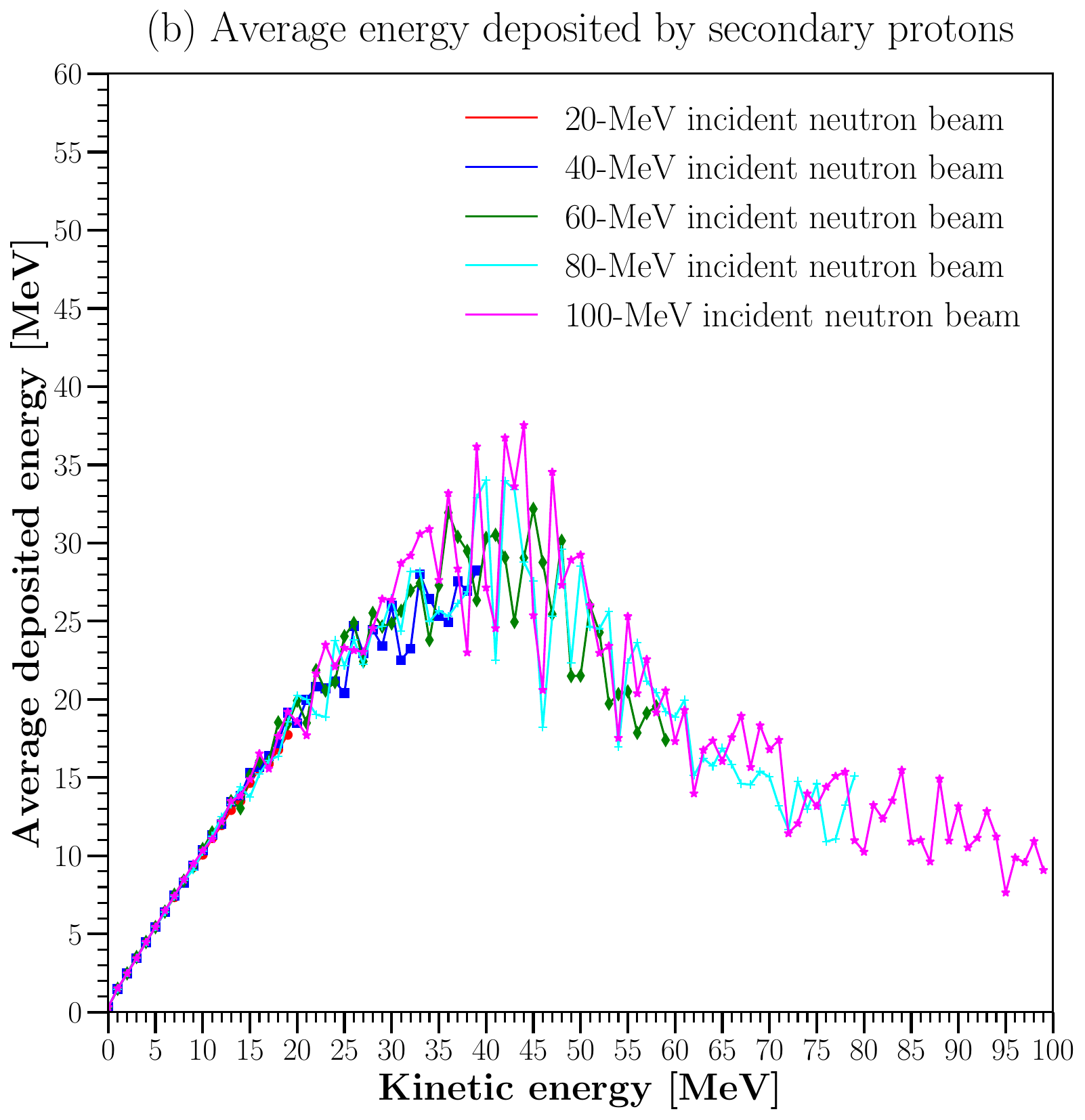}
\caption{Energy deposition of secondary protons within the present SONTRAC detector irradiated by fast neutron beam: (a) distribution of energy deposition by secondary protons with a step length of 0.1 MeV and (b) average deposited energy by secondary protons with a bin size of 1 MeV.}
\label{Proton_energy_deposition}
\end{center}
\end{figure}

One of the crucial parameters in the detection of the solar neutrons is the range of the secondary protons since the SONTRAC concept is hinged on the tracking of the secondary protons. The range of the secondary protons is directly dependent on the kinetic energy of the secondary protons; however, it should be noted that the initial generation point of a secondary proton as well as its momentum direction also notably influences the numerical values of the secondary proton range. By reminding that these two parameters of a secondary proton are randomly sampled at the interaction location of the incident fast neutrons within the detector parts, the range of a sufficiently high energy proton might still be short. The range of the secondary protons is computed through the spatial coordinates of the generation point together with the last location for the secondary protons within the constituents of the current SONTRAC detector that includes the fiber core, the first clad, and the second clad. The distribution of the secondary proton ranges by using a set of the initial neutron energies between 20 and 100 MeV is exhibited in Fig.~\ref{Proton_range}(a) with a step length of 0.1 cm, and it is demonstrated that the proton range rises if the kinetic energy of the incident neutrons increases because the kinetic energy of the generated protons depends on the kinetic energy of the incident neutrons. In contrast, in every neutron beam, a greater part of the secondary protons stop within a short range as might be comprehended from Fig.~\ref{Proton_range}(a). While the 20-MeV neutron beam along with the 40-MeV neutron beam leads to the distinct histograms, the rest of the fast neutron beams with the kinetic energies of 60, 80, and 100 MeV result in strongly similar outcomes. In Fig.~\ref{Proton_range}(b), the average proton range is also calculated for each energy bin where the bin step length is 1 MeV, and it is shown that the secondary proton range increases as a function of the initial kinetic energy. 
In the latter step, the energy deposition by the secondary protons is furthermore explored, and the distribution of the deposited energy is depicted in Fig.~\ref{Proton_energy_deposition}(a) by using a step length of 0.1 MeV. 
From Fig.~\ref{Proton_energy_deposition}(a), it is observed that the maximum energy deposited by the secondary protons in the cases of the fast neutron beams with the kinetic energies of 60, 80, and 100 MeV is around 50 MeV, while the secondary protons generated by either the 20-MeV neutron beam or the 40-MeV neutron beam are able to deposit their entire energies within the current SONTRAC detector depending on the location of emission. This outcome is also validated by computing the average deposited energy for each energy bin with a bin size of 1 MeV as described in Fig.~\ref{Proton_energy_deposition}(b), and it is revealed that the secondary protons up to 20 MeV in every fast neutron beam completely lose their energies within the detector sections.
\subsection{Generation of optical photons}
\begin{figure}[H]
\begin{center}
\includegraphics[width=8cm]{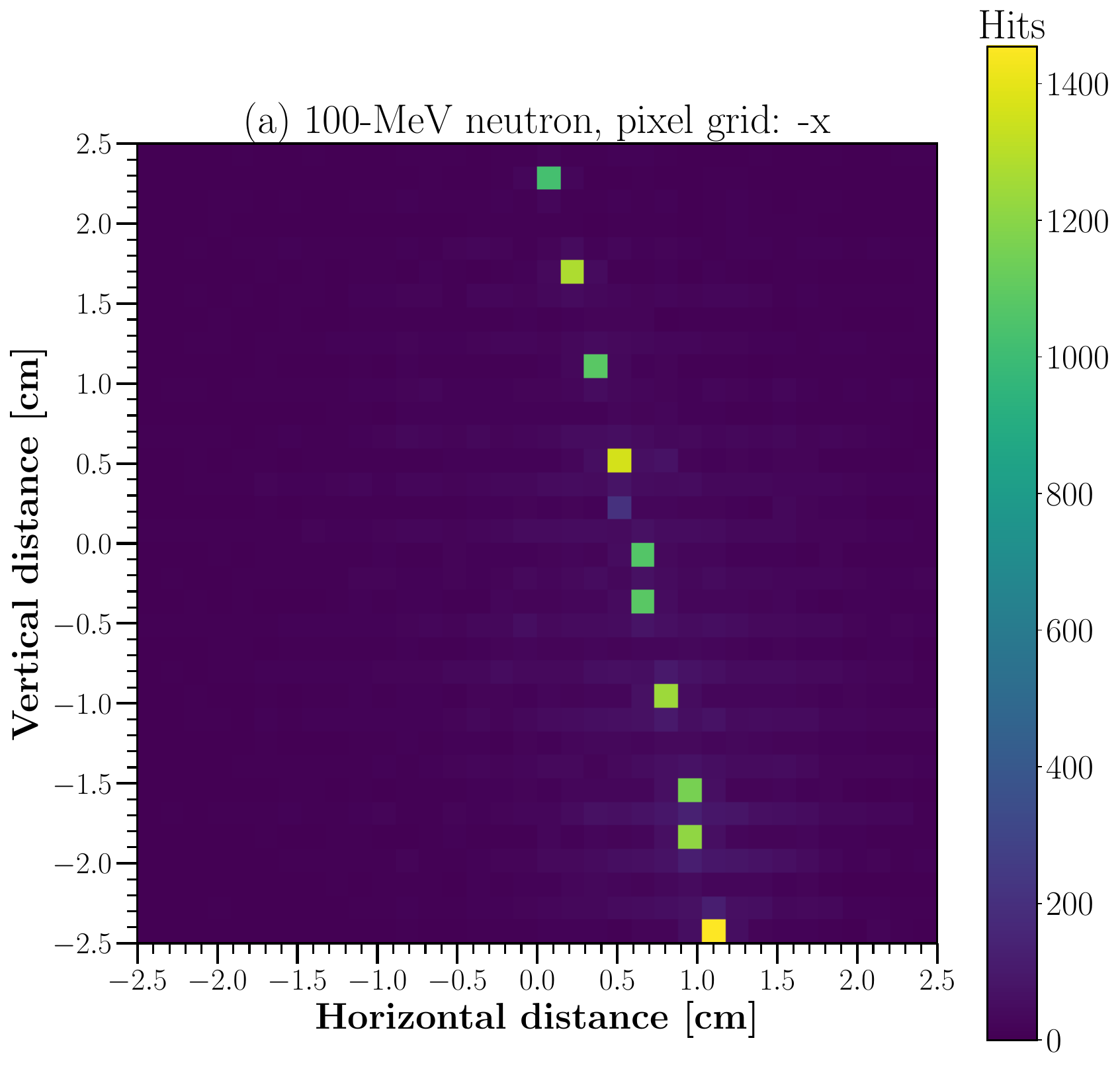}
\includegraphics[width=8cm]{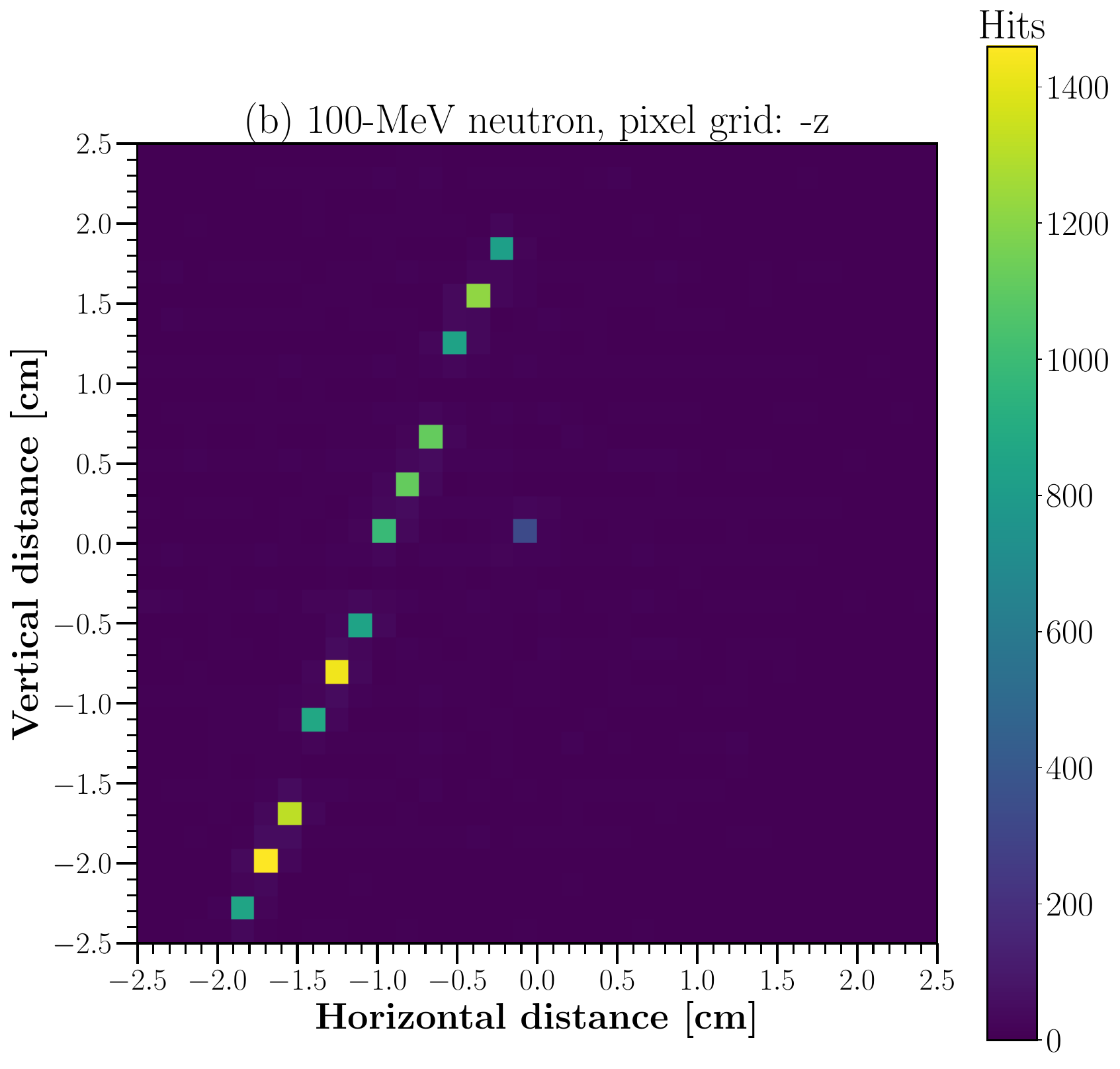}
\includegraphics[width=8cm]{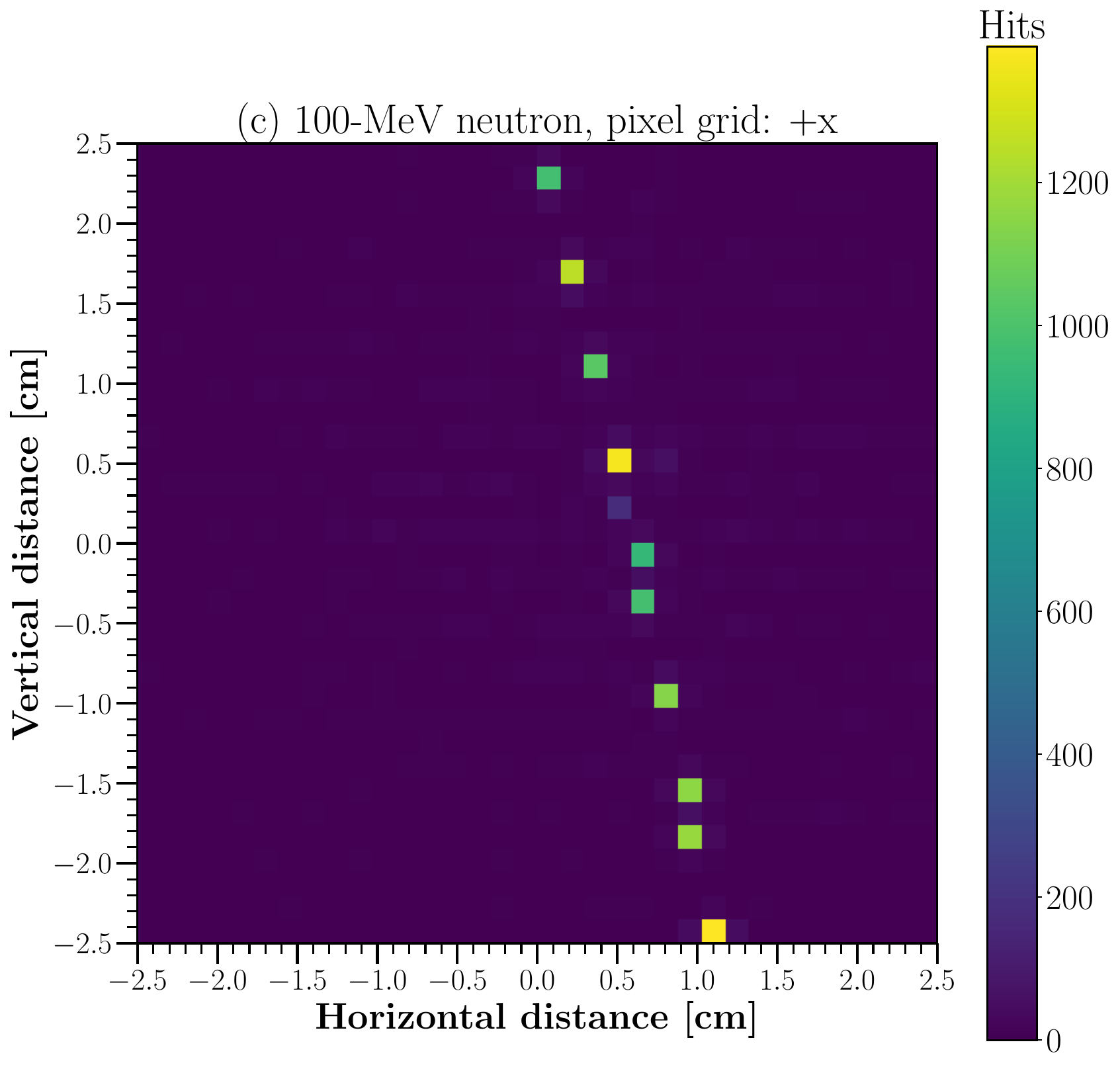}
\includegraphics[width=8cm]{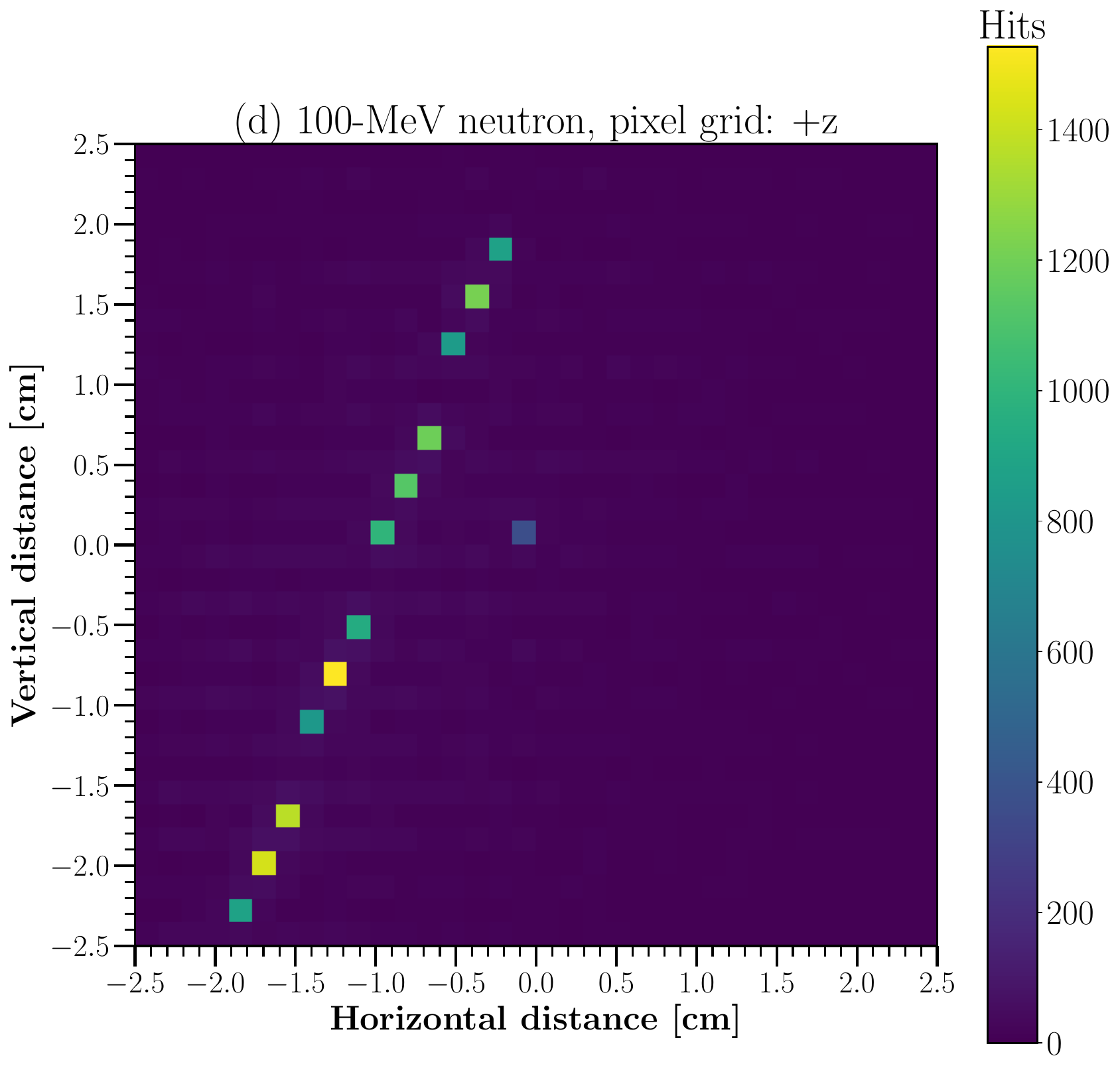}
\caption{Hits of optical photons on pixel grid detectors due to a secondary proton of 80 MeV from a neutron beam of 100 MeV: (a) -x position, (b) -z position, (c) +x position, and (d) +z position. }
\label{Optical_photons_100_MeV}
\end{center}
\end{figure}
As initially explained, the generated secondary particles lose their kinetic energies in the course of the propagation through out the scintillation fibers, and the deposited energy in these scintillation fibers leads to the emission of the optical photons. In this section, it is aimed at demonstrating the trajectories of the secondary charged particles via the release of the optical photons from the scintillation fibers as a consequence of the energy deposition, i.e. particularly the secondary protons since the secondary alphas as well as the secondary ions frequently are unable to penetrate multiple fibers. To attain this goal, a 34$\times$34 pixel grid detector that points out each Kuraray Y11-200(M) fiber is inserted on each side of the fiber bundle. For the sake of simplicity,  a similar planar vertical neutron beam of 100 MeV that consists of 100 neutrons is injected. It is worth stating that the GEANT4 simulations that incorporate the optical photons are significantly slower compared to the cases without the optical photons. The optical photons on each pixel of 1.36$\times$1.36 mm$^{2}$ are processed, and the number of hits by the optical photons is counted. Four pixel grids are situated along the x-direction as well as the z-direction including their opposite axes around the present SONTRAC detector. Fig.~\ref{Optical_photons_100_MeV} depicts the path of a 80-MeV traversing secondary proton that is generated at the upper part of the present fiber bundle. By rementioning that the maximum deposited energy by the secondary protons is around 50 MeV as seen in Fig.~\ref{Proton_energy_deposition}(a), the secondary protons with the kinetic energies above 60 MeV almost assure to cross the current SONTRAC detector from any location of generation. However, one should note that a notable number of the low-energetic protons might be also generated even if the kinetic energy of the incident neutrons is above 60 MeV as demonstrated in Figs.~\ref{Secondary_energy_spectrum}(a). Fig.~\ref{Optical_photons_100_MeV} confirms the presence of the secondary charged particles due to the fast neutron irradiation via the hits on the pixel photo-collectors, and it is observed that the pixel grid detector provides substantial information about the range as well as the energy deposition for the secondary charged particles generated from the present SONTRAC detector, thereby bringing forth the possibility of the track reconstruction for the secondary protons by combining the pixel grid detectors especially in the case of the kinetic energies above 60 MeV.
\section{Conclusion}
\label{Conclusion}
In this study, the secondary particle emission owing to the incident fast neutrons from a SONTRAC detector based on Kuraray Y11-200(M) scintillation fibers is investigated by means of GEANT4 simulations. It is shown that the fast neutron irradiation leads to the generation of a relatively low number of secondary protons that might be detected through the energy deposition in the plastic scintillators, thereby providing a possibility to indirectly track the incident fast neutrons with a low detection efficiency. According to the present computations, the secondary protons with the kinetic energies above 60 MeV are capable of traveling across the current SONTRAC detector, whereas the secondary protons, the kinetic energies of which are below 20 MeV, lose their entire kinetic energies and stop within a short range along the current fiber bundle. At last, the collection of the optical photons from the scintillation fibers due to the energy deposition of the secondary particles is also exhibited by the aid of pixel grid detectors that surround the current fiber bundle, and it is observed that it is possible to perform the track reconstruction of the secondary protons in accordance with the hit numbers on the specific pixels.
\bibliographystyle{elsarticle-num}
\bibliography{SONTRAC} 
\end{document}